\documentclass[prl,aps,twocolumn,tightenlines,notitlepage,nofootinbib,preprintnumbers,letterpaper,superscriptaddress]{revtex4-1} 
\pdfoutput=1
\usepackage{epsfig}
\usepackage{amsfonts}
\usepackage{amsmath}
\usepackage{graphicx}
\usepackage{color}
\usepackage{amssymb,amsmath,hyperref,slashed,color,graphicx,bm}
\usepackage[T1]{fontenc}
\usepackage{relsize}

\hypersetup{colorlinks,citecolor= nicegreen,linkcolor= nicered}
\definecolor{nicered}{rgb}{0.7,0.1,0.1}
\definecolor{nicegreen}{rgb}{0.1,0.5,0.1}

\newcommand{\beq}{\begin{eqnarray}}
\newcommand{\eeq}{\end{eqnarray}}

\def\babar{\mbox{\slshape B\kern-0.1em{\smaller A}\kern-0.1em B\kern-0.1em{\smaller A\kern-0.2em R}}}

\begin{document}

\title{Probing the Dark Sector with Dark Matter Bound States}
\author{Haipeng An}
\affiliation{Walter Burke Institute for Theoretical Physics,
California Institute of Technology, Pasadena, CA 91125}
\author{Bertrand Echenard}
\affiliation{California Institute of Technology, Pasadena, California 91125}
\author{Maxim Pospelov}
\affiliation{Department of Physics and Astronomy, University of Victoria,
  Victoria, BC, V8P 1A1 Canada}
\affiliation{Perimeter Institute for Theoretical Physics, Waterloo,
ON, N2L 2Y5, Canada}
\author{Yue Zhang}
\affiliation{Walter Burke Institute for Theoretical Physics,
California Institute of Technology, Pasadena, CA 91125}

\date{\today}

\begin{abstract}
A model of dark sector where $O({\rm few~GeV})$ mass dark matter particles $\chi$ 
couple to
 a lighter dark force mediator $V$, $m_V \ll m_\chi$, is motivated by the recently discovered mismatch between simulated and observed shapes of galactic haloes. Such models, in general, provide a challenge for direct detection efforts and collider searches. We show that for a large range of coupling constants and masses, the production and decay of the bound states of $\chi$, such as $0^{-+}$ and $1^{--}$ states, $\eta_D$ and $ \Upsilon_D$, is an important search channel. We show that $e^+e^-\to \eta_D +V$ or $\Upsilon_D +\gamma$ production at $B$-factories for $\alpha_D > 0.1$ is sufficiently strong to result in multiple pairs of charged leptons and pions via $\eta_D\to 2V \to 2(l^+l^-)$ and $\Upsilon_D\to 3V \to 3(l^+l^-)$ $(l=e,\mu,\pi)$. The absence of such final states in the existing searches performed at \babar\ and Belle sets new constraints on the parameter space of the model. We also show that a search for multiple bremsstrahlung of dark force mediators, $e^+e^-\to \chi\bar\chi+nV$, resulting in missing energy and multiple leptons, will further improve the sensitivity to self-interacting dark matter. 


\end{abstract}

\preprint{CALT-TH-2015-051}
\maketitle

\noindent{\sf\bfseries Introduction.}
Identifying dark matter is an open question of central importance in particle physics and cosmology.
In recent years, the paradigm of weakly interacting dark matter supplied by a new force in the dark sector 
came to prominence~\cite{Pospelov:2007mp, ArkaniHamed:2008qn}, motivated by a variety of 
unexplained astrophysical signatures.  It was later shown~\cite{Tulin:2013teo, Kaplinghat:2015aga} that this model
provides a straightforward realization of self-interaction dark matter~\cite{Spergel:1999mh}, 
which helps to alleviate tensions between observed and simulated shapes of dark matter haloes (see, {\em e.g.} \cite{BoylanKolchin:2011de}).

It is of great phenomenological interests to check whether such a dark force could be probed in laboratories.
The simplest way for dark matter to interact with the standard model (SM) sector is 
through a vector or scalar mediators coupled to the SM fields via the kinetic mixing or the Higgs portals.
For dark matter heavier than 4-5\,GeV, direct detection experiments provide the strongest constraints on such models.
High-energy collider probes typically require more effective production channels~\cite{Shepherd:2009sa, Autran:2015mfa, Bai:2015nfa, Buschmann:2015awa, tsai}.
For dark matter lighter than 4-5\,GeV, the limits from direct detection experiments arise from electron recoil~\cite{Essig:2012yx} and are much weaker.
In this mass range, strong CMB constraints on dark matter annihilation~\cite{Ade:2015xua,Slatyer:2015jla} naturally point to particle-antiparticle asymmetry in the dark 
sector. Constituents of such a dark sector, light dark matter and a light mediator, can be searched for in meson decays~\cite{Fayet:2006sp}, fixed 
target experiments~\cite{Izaguirre:2014bca}, mono-photon events at colliders~\cite{Essig:2013vha}, or via the production/scattering 
sequence in proton~\cite{Batell:2009di} and electron~\cite{Izaguirre:2013uxa} beam dump experiments, 
or perhaps via new galactic substructures and minihalos~\cite{Zhang:2015era}.
Most of the existing searches of light particles~\cite{Essig:2013lka} are insensitive to dark matter with $m_\chi> m_{\rm mediator}$, and 
therefore would not be able to establish  any candidate signal as coming specifically from the dark force carrier.

In this {\it Letter}, we show that the presence of self-interacting dark matter within the kinematic 
reach of existing colliders provides opportunities for the new search channels. 
We outline such possibilities in the minimal setup where the dark force carrier also mediates the interaction between dark matter and the SM particles.
A light mediator gives an attractive force between $\chi$ and $\bar\chi$ particles, leading to the formation of bound states,
which can be produced on-shell at colliders~\footnote{Weakly coupled dark matter bound states have been studied in various 
contexts~\cite{Pospelov:2008jd,MarchRussell:2008tu, Kaplan:2009de, Braaten:2013tza, Pearce:2013ola, Wise:2014jva, Wise:2014ola}.}.
In addition,  the production of continuum $\chi\bar\chi$ leads to final state radiation (FSR) of light mediators.
Both channels typically result in a striking multi-lepton final state, that can be searched for at $B$-factories and fixed target 
experiments. It is well known that heavy flavor mesons and heavy quarkonia were instrumental for uncovering a wealth of information about the SM. Similarly, should 
a dark force exist, the aforementioned channels may allow for genuine tests of the detailed content of the dark sector.

\smallskip
\noindent{\sf\bfseries Dark matter bound states production.}
We illustrate these ideas in the well-studied example of the vector mediator model. The Lagrangian for dark matter and dark photon is
\begin{eqnarray}\label{Lvector}
\mathcal{L} &=& \mathcal{L}_{\rm SM} + \bar\chi i\gamma^\mu(\partial_\mu - i g_D V_\mu) \chi - m_\chi \bar \chi \chi \nonumber \\
&& -\frac{1}{4} V_{\mu\nu} V^{\mu\nu} - \frac{\kappa}{2} F_{\mu\nu} V^{\mu\nu} + \frac{1}{2} m^2_{V} V_\mu V^\mu \ ,
\end{eqnarray}
where $\kappa$ is the kinetic mixing between the photon and the vector field $V$.
The dark matter particle $\chi$ is a Dirac fermion, neutral under the SM gauge group, but charged under the dark $U(1)_D$ interaction
that has a new vector particle $V_\mu$ (sometimes called a "dark photon") as a force carrier. It is assumed that the correct cosmological 
abundance of dark matter is controlled by particle-antiparticle asymmetry in the dark sector. 
(Other well-motivated realizations of self-interacting dark matter based on a new strongly interacting sector 
would also typically require the existence of dark photons \cite{Hochberg:2014dra}.)

\begin{figure}[h!]
\centerline{\includegraphics[width=0.51\columnwidth]{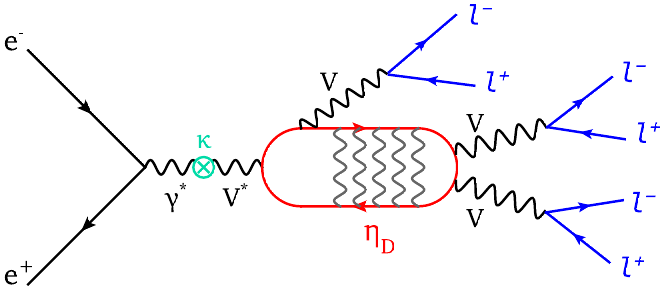}\hspace{-0.1cm}
\includegraphics[width=0.51\columnwidth]{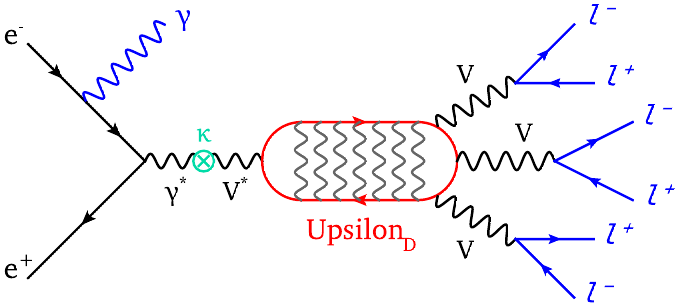}}
\caption{Diagram for $\eta_D$ and $\Upsilon_D$ production and decay at $B$-factories.}\label{Fig:Feynman-BABAR}
\end{figure}

As discussed in the introduction, sufficiently strong dark interaction strength and light dark photon will result in the formation of dark matter particles ($\chi\bar\chi$). 
The two lowest  $(1S)$ bound states, $^1S_0$ ($J^{PC}=0^{-+}$) and $^3S_1$ ($J^{PC}=1^{--}$), will be called $\eta_D$ and $\Upsilon_D$, respectively.
The condition for their existence has been determined numerically~\cite{rogers1970}~\footnote{It 
is known that too large $\alpha_D$ would run to the Landau pole very quickly at higher scale~\cite{Davoudiasl:2015hxa}. Hereafter, we focus on $\alpha_D\leq0.5$, 
and work with leading-order results in $\alpha_D$.}, $1.68 m_V< \alpha_D m_\chi$, with $\alpha_D=g_D^2/(4\pi)$. Their quantum numbers suggest the following production mechanisms at colliders:
\begin{equation}
e^+e^-\to \eta_D +V;\ \ \ e^+e^-\to \Upsilon_D +\gamma;\ \ \ p+p \to \Upsilon_D + X
\label{modes}
\end{equation}
The last process represents the direct production of $\Upsilon_D$ from $q\bar q$ fusion. 
All production processes are mediated by a mixed $\gamma-V$ propagator, as shown in Fig.~\ref{Fig:Feynman-BABAR}.

In order to obtain the rate for the first process in (\ref{modes}), we calculate the amplitude of $e^+e^-\to \chi \bar \chi V$ 
with $\chi, \bar\chi$ having the same four momentum $p$ (with $p^2=m_\chi^2$), and apply the projection operator, 
\begin{eqnarray}
\Pi_\eta=\sqrt{\frac{1}{32\pi m_\chi^3}}R_{\eta_D}(0)(\not\!{p}+m_\chi)\gamma_5 (\not\!{p}-m_\chi) \ ,
\end{eqnarray}
to select the $\eta_D$ bound state~\cite{Petrelli:1997ge}. We find a leading-order differential cross section:
\begin{eqnarray}\label{sigma1}
\frac{d \sigma_{e^+e^-\to \eta_D V}}{d \cos\theta} = \frac{4\pi\alpha \alpha_D^2 \kappa^2 [R_{\eta_D}(0)]^2 (1+\cos^2\theta)}{m_\chi s^{3/2}(s-4m_\chi^2+m_V^2)^2} |{\bf p}|^3\,, \ \ \ 
\end{eqnarray}
where $\theta$ is the angle between $\eta_D$ and the initial $e^-$ in the center-of-mass (CM) frame, and $|{\bf p}|$ is the spatial momentum of $\eta_D$, $|{\bf p}|=\sqrt{[s-(2m_\chi+m_V)^2][s-(2m_\chi-m_V)^2]}/(2\sqrt{s})$. We neglect the binding energy for $\eta_D$, and set $m_{\eta_D}\simeq 2 m_\chi$.

The value of $R_{\eta_D}(0)$ can be analytically estimated using the Hulth\'en potential $V(r) = - {\alpha_D \delta e^{-\delta r}}/({1-e^{-\delta r}})$ 
with $\delta=({\pi^2}/{6}) m_V$, known as a good approximation of the Yukawa potential $V(r)=-{\alpha_D e^{-m_V r}}/{r}$~\cite{lam}. In that case, 
$R_{\eta_D}(0) = (4-\delta^2 a_0^2)^{1/2} a_0^{-3/2}$, where $a_0 = {2}/({\alpha_D m_\chi})$. 


The scalar bound state $\eta_D$ dominantly decays into two dark photons, each subsequently decaying into a pair of SM particles via kinetic mixing. These decays are all prompt 
for the relevant region of parameter space. The above decay chain eventually results in the final states containing six charged tracks, which can be electrons, muons or pions, 
depending on the dark photon mass. 

We turn to the calculation of $\Upsilon_D$ production via initial state radiation (Fig.~\ref{Fig:Feynman-BABAR}).
In the $\Upsilon_D$ rest frame, the non-relativistic expansion can be used, taking the dark matter field in the form:
$\chi = e^{i m_\chi t} \left[ 
\xi, 
{\sigma\cdot {\bf p}}/({2m_\chi}) \xi \right]^T
+
e^{-i m_\chi t}\left[ 
{\sigma\cdot {\bf p}}/({2m_\chi}) \zeta,
\zeta \right]^T$, 
where $\xi$, $\zeta$ are the 2-spinor annihilation (creation) operators for particle (antiparticle). We use the relation between matrix element and wave function~\cite{Bodwin:1994jh},
\begin{eqnarray}
\langle0| \zeta^\dagger \sigma^\mu \xi | \Upsilon_D \rangle = \sqrt{\frac{1}{2\pi}} R_{\Upsilon_D}(0)\, \varepsilon_{\Upsilon_D}^\mu \ ,
\end{eqnarray}
where $\varepsilon_{\Upsilon_D}^\mu$ is the polarization vector of $\Upsilon_D$ and $R_{\Upsilon_D}(0)\simeq R_{\eta_D}(0)$ is the radial wave function at origin.
Taking into account the kinetic mixing between dark photon and the photon, we derive the effective kinetic mixing term between $\Upsilon_D$ and the photon,
\begin{eqnarray}
\mathcal{L}_{\rm eff} = - \frac{1}{2} \kappa \kappa_D F_{\mu\nu} \Upsilon_D^{\mu\nu}, \ \ \  \kappa_D = \sqrt{\frac{\alpha_D}{2 m_\chi^3}} R_{\Upsilon_D}(0) \ .
\end{eqnarray}
In the limit $m_V\ll\alpha_D m_\chi$, the term $\kappa_D$ reduces to $\kappa_D=\alpha_D^2/2$. We obtain a differential cross section: 
\begin{eqnarray}\label{sigma2}
&&\frac{d \sigma_{e^+e^-\to \gamma\Upsilon_D}}{d\cos\theta} \simeq \frac{2\pi\alpha^2 \kappa^2 \kappa_D^2}{s} \left( 1- \frac{4m_\chi^2}{s} \right) \nonumber \\
&&\hspace{0.4cm}\times \left[ \frac{8s^2(s^2+16m_\chi^4) \sin^2\theta}{(s-4m_\chi)^2 \left(s+4m_e^2 - (s-4m_e^2)\cos2\theta\right)^2} - 1 \right], \ \ \ \ \
\end{eqnarray}
where $\theta$ is the the angle between $\gamma$ and the initial $e^-$ in the CM frame. 
In the denominator, the electron mass must be retained in order to regularize the $\theta$ integral, as for $m_e=0$ the cross section is divergent in the forward direction~\cite{Fayet:2007ua}.

Compared to the $e^+e^-\to \eta_D V$ process, the $e^+e^-\to \gamma\Upsilon_D$ cross section is suppressed by a factor $\alpha/\alpha_D$, although the latter contains a 
logarithmic enhancement from the angular integral.
Moreover, the cross-section $e^+e^-\to \eta_D V$ contains an additional $m_\chi^2/s$ factor, which brings additional suppression of lighter dark matter. 
For $\alpha_D\gtrsim0.1$ and $m_\chi\sim \sqrt{s}$, the two processes have similar cross-sections, and we will combine them to set the limit on this model.

The $\Upsilon_D$ particle will subsequently decay into three dark photons.
Similarly to derivation of Eq.~(\ref{sigma1}), we calculate the differential decay rate of $\Upsilon_D$ into three massive dark photons,
\begin{eqnarray}\label{decayrate1}
&&\frac{d \Gamma(\Upsilon_D \to 3 V)}{d x_1 d x_2} = \frac{2\alpha_D^3 \left[R_{\Upsilon_D}(0)\right]^2}{9\pi m_\chi^2} \nonumber\\
&&\hspace{0.6cm}\times
\frac{39 x^8 + 4 x^6 F_6 -16 x^4 F_4 + 32 x^2 F_2 + 256 F_0}{(x^2-2x_1)^2(x^2-2x_2)^2(x^2+2(x_1+x_2-2))^2} \ , \ \hspace{0.3cm}
\end{eqnarray}
where $x_{1,2} = E_{1,2}/m_\chi$, $x= m_V/m_\chi$, and 
\begin{eqnarray}
F_6 &=& x_1^2 + (x_1+x_2)(x_2-2) - 30, \nonumber\\
F_4 &=& (x_1^2 + x_1 x_2-2 x_1)(3x_2-10) - 10 x_2(x_2-2)-21, \nonumber \\
F_2 &=& x_1^4 + 2 x_1^3 (x_2-2) + x_1^2 (x_2 (3 x_2 -22)+28) \nonumber\\
           &  & + 2 x_1 (x_2-2) (x_2(x_2-9)+12) \nonumber\\
           &  & + x_2 (x_2-2)(x_2(x_2-2)+24) + 24, \nonumber \\
F_0 &=& x_1^4 + 2 x_1^3(x_2-2) + x_1^2 (3x_2(x_2-3)+7)  \nonumber\\
           &  & + x_1 (x_2-1)(x_2-2)(2x_2-3)  \nonumber\\
           &  & + (x_2-1)^2 (x_2(x_2-2)+2) \ .         
\end{eqnarray}
When $x_1, x_2$ are fixed, the relative angles between the dark photons are also fixed in the rest frame of $\Upsilon_D$.
Their ranges are $x \leq x_1 \leq 1-({3}/{4}) x^2$, and
$(x_2)^{\rm max}_{\rm min} = \pm \sqrt{{(4-3x^2-4x_1)(x_1^2 -x^2)}/({4+ x^2 - 4 x_1})}/2 + (2-x_1)/2$.
This channel eventually results in the final states containing 3 pairs of electrons, muons or pions, and one photon.

\begin{figure*}[t!]
\centerline{\includegraphics[width=1.016\columnwidth]{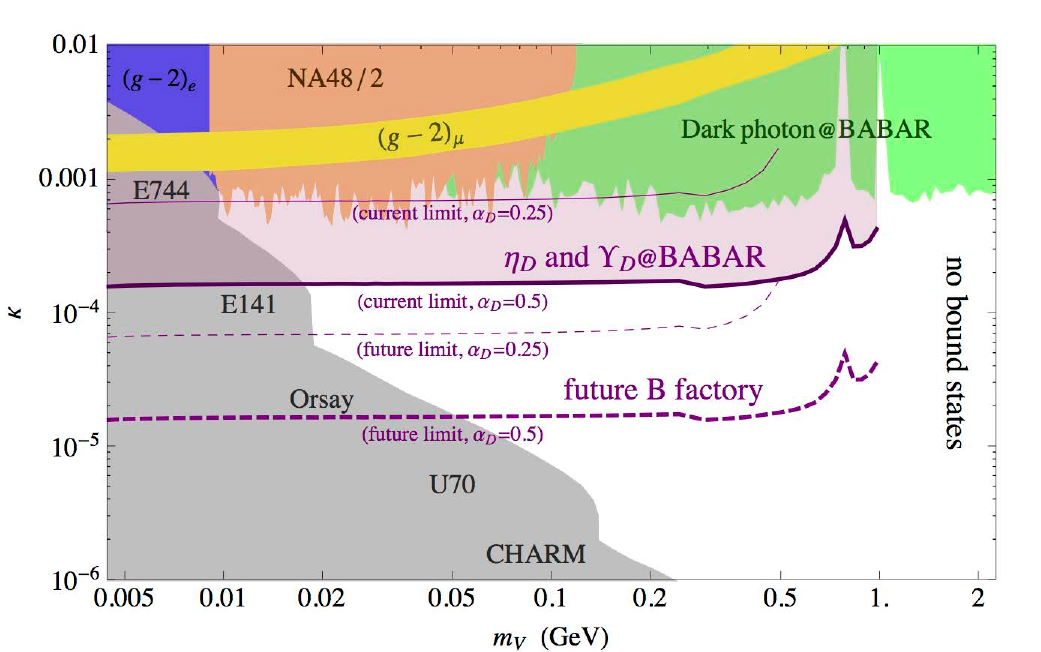}
\includegraphics[width=0.95\columnwidth]{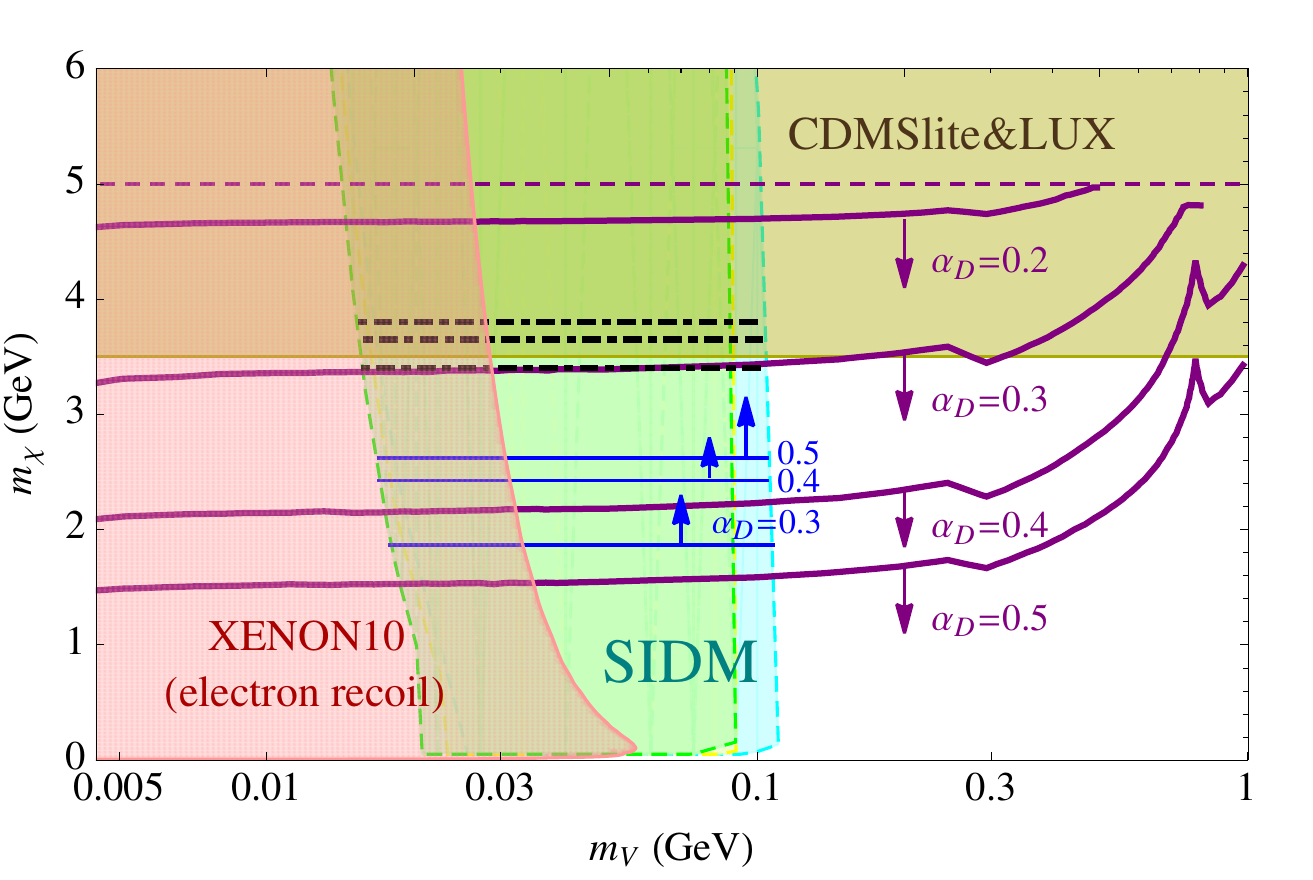}}
\caption{{\it Left}: Constraint on the dark photon parameter space from the \babar\ dark Higgsstrahlung searches, adapted to the production and decay of
dark bound states $\eta_D$ and $\Upsilon_D$. The solid purple curve corresponds to the  current \babar\ limit for the parameters $\alpha_D=0.5$, $m_\chi=3.5$\,GeV. 
The dashed purple curve shows the future reach of B-factories.
{\it Right}: Current constraints on the $m_\chi-m_V$ plane for the SIDM scenario are shown with $\kappa^2 = 10^{-7}$ and different values of $\alpha_D$. 
The green (blue) region is favored for SIDM solving the galactic small-scale structure problems~\cite{Tulin:2013teo} for $\alpha_D=0.3\, (0.5)$. 
The combined constraints via the $e^+e^-\to (\eta_DV,\ \Upsilon_D) \to 3V$ channels are shown in thick purple curves,
and the constraints via the $e^+e^-\to \chi\bar\chi + 3V$ channel are shown in thin blue curves. Allowed regions are in the arrow direction.
Assuming no SM background, the constraints via the $e^+e^-\to \chi\bar\chi + 2V$ channel are shown in dot-dashed black curves for $\alpha_D=0.3, 0.4, 0.5$ (bottom-up).
The brown region is excluded by CDMSlite~\cite{Agnese:2013jaa} and LUX~\cite{Akerib:2013tjd}. 
The region $m_V\lesssim30$\,MeV is ruled out by the XENON10 electron recoil analysis~\cite{Essig:2012yx} for $\alpha_D=0.3$.
}\label{Fig:DP-BABAR}
\end{figure*}

To estimate the limit from searches at $B$-factories, we simulate events according to (\ref{sigma1}) and (\ref{sigma2}), and apply the kinematic constraints used for dark Higgsstrahlung 
searches~\cite{Batell:2009yf,Lees:2012ra,TheBelle:2015mwa}. We select events containing six charged tracks (made of $e$, $\mu$ or $\pi$), 
excluding the $6\pi$ final states due to the presence of larger SM backgrounds. For each track, we require $p_T>150\,$MeV, and $-0.95<\cos\theta_{\rm cm} < 0.85$ 
in the CM frame. We include a 95\% efficiency for reconstructing each of the charged track. 
For the $\eta_D$ channel, we require the invariant mass of the six charged tracks to be close to the CM energy.
For $\Upsilon_D$ channel, we do not require to find the photon, but impose the condition that the missing mass recoiling against the six tracks is around zero. 
We assume negligible SM background, similarly to the dark Higgsstrahlung searches~\cite{Batell:2009yf,Lees:2012ra,TheBelle:2015mwa}. 
In the left plot of Fig.~\ref{Fig:DP-BABAR}, we present the 90\% C.L. exclusion in the dark photon parameter space based the existing \babar\ luminosity 
of 516\,fb$^{-1}$~\cite{Lees:2012ra}. We expect the current Belle data to give a similar limit.
If dark bound states exist, the limit on the kinetic mixing is more than one order of magnitude stronger than 
the direct dark photon search via di-lepton resonance at \babar\ \cite{Lees:2014xha} and the pion decay search at NA48/2~\cite{Batley:2015lha}. 
We also show the expected sensitivity of future $B$-factories (Belle-II) assuming 100 times more luminosity and similar search strategy.

These results provide useful constraints on self-interacting dark matter (SIDM) scenarios. In the right plot of Fig.~\ref{Fig:DP-BABAR}, the green region is favored for SIDM models solving the small-scale structure problem, which satisfies the condition
$0.1\,{\rm cm^2/g}\leq\langle \sigma_T\rangle/m_\chi\leq10\,{\rm cm^2/g}$~\cite{Tulin:2013teo}.
For this parameter space, the $s$-partial wave gives the dominant contribution to the dark matter elastic scattering cross section.
The purple curves show the current \babar\ 90\% C.L. exclusion contours for the SIDM model. 
Compared to direct detection experiments, dark bound states at \babar\ further constrain the allowed dark matter mass down to sub-GeV, if $\alpha_D$ is sufficiently 
large.

In the case of scalar dark matter charged under $U(1)_D$, the ground state $\chi_D$ formed by a pair $\chi \chi^*$ has quantum numbers $^1S_0$ ($0^{++}$)~\cite{book}, 
and will be produced in the similar process as $\eta_D$ in (\ref{modes}). On the other hand, the counterpart of $\Upsilon_D$ is a $p$-wave state, and its 
production rate is further suppressed by the derivative of its wave function at the origin. Therefore, we expect slightly weaker bounds on scalar dark matter 
compared to the fermion case.

\smallskip
\noindent{\sf\bfseries Multi-mediator final state radiation.}
Smaller values of $\alpha_D$ or larger $m_V/m_\chi$ ratios may prevent the existence of $\chi\bar\chi$  bound states. 
In that case, mediator states can still be produced through the FSR process $e^+e^-\to \chi\bar \chi +n V $. 
(One could also study this process in high-energy proton collisions \cite{Buschmann:2015awa}, should a new efficient channel for 
$\chi\bar\chi$ production exist.)
The FSR dark photons further decay into pairs of charged SM particles. Therefore, the typical signal consists of multiple charged tracks plus missing energy,
taken away by the $\chi\bar\chi$ pair. 
The \babar\ experiment did not trigger on two charged leptons due to overwhelming QED backgrounds. 
The channel with four charged leptons plus missing energy is, however, quite promising, and we suggest to perform a 
corresponding search at both \babar\ and Belle. 
The dominant SM backgrounds for the $4l$~+~missing energy signature may come from the $\tau^+\tau^-l^+l^-$ final states, and one would expect 
over $10^4$ such events at \babar. If however, the two invariant mass $m_{l^+l^-}=m_V$ conditions are imposed, this background can be considerably reduced. 
With the assumption of negligible background, the whole low mass dark matter window for the SIDM can be potentially ruled out for $\alpha_D\gtrsim0.3$, as shown 
in 
black
dot-dashed curves of Fig.~\ref{Fig:DP-BABAR}. 
Six charged lepton final states, similar to the case of the bound state study, have been searched for, and we generate 
$e^+e^-\to \chi\bar \chi+3V$ events using MadGraph5~\cite{Alwall:2011uj}.
With the same kinematic requirements described in the previous section, the lower bounds on $m_\chi$ in the region favored by the SIDM model are shown by the thin blue curves in Fig.~\ref{Fig:DP-BABAR} for several choices of $\alpha_D$. 
For this channel, we only show the constraint in the region of interests to the SIDM scenario. For smaller $m_V$, the leading order simulation becomes less accurate due to the large logarithms from the soft dark photons. Moreover, this region has already been excluded by the direct detection experiment with electron recoils~\cite{Essig:2012yx}. 
For  $m_V>{\rm few}~100$ MeV the sensitivity is expected to worsen due to a shrinking phases space. 

The search for FSR production of dark photons by dark matter pair-production has additional kinematic limitations. 
The phase space for producing energetic charged leptons becomes smaller for larger $m_\chi$, resulting 
in softer final state leptons. This feature can be read from Fig.~\ref{Fig:DP-BABAR}, as for $m_\chi \gtrsim$ 2.5 GeV, 
producing charged leptons energetic enough to pass the cuts becomes difficult. 
As a result, the potential lower bound on $m_\chi$ does not change very much with the increase of $\alpha_D $. 
On the other hand, the production and decay of dark bound states $\Upsilon_D$ and  $\eta_D$ create more energetic leptons for larger $m_\chi$. 
Therefore, the two search strategies are complementary to each other. 

\smallskip
\noindent{\sf\bfseries Hadronic probes of dark sector.} 
Fixed target experiments with proton beams can also be used to probe a dark sector. For realistic energies of available proton beams, the most 
important production channel is from the quark-anti-quark fusion, $q \bar q \to \Upsilon_D$. 
Generalizing calculations of \cite{deNiverville:2012ij}, the production cross section is given by
\begin{eqnarray}
\label{pp}
\sigma_{pp(n)\to \Upsilon_D}= \frac{ 4 \pi^2 \alpha \kappa^2 \kappa_D^2}{s}\sum_qQ_q^2\int_{\tau}^1\frac{dx}{x} \hspace{2cm}  \nonumber \\
\times\left[ f_{q/p}(x) f_{\bar q /p(n)}\left(\frac{\tau}{x}  \right) + f_{\bar q/p}(x) f_{ q /p(n)}\left(\frac{\tau}{x}\right)\right], \hspace{0.5cm}
\end{eqnarray}
where $\tau = m_V^2/s$, $f_{q/p(n)}$ and $f_{\bar q/p(n)}$ are the relevant structure functions for this process, and $Q_q$ is the quark charge in units of $e$.
Unlike $B$-factories, only muonic decays of dark bound states, such as $\Upsilon_D\to 3V\to 3(\mu^+\mu^-)$, constitute a useful signature, 
as backgrounds in other channels are likely to be too large. The multi-dark photon FSR channels can also be relevant for the proton beam experiments.

Among the possible candidates of proton-on-target experiments, we focus our discussion on SeaQuest~\cite{seaquest, Gardner:2015wea} and the planned SHiP~\cite{Alekhin:2015byh} facilities.
Note that only a fixed target mode of operation, rather than a beam dump mode that would try to remove prompt muons, 
is suitable for the search of $\Upsilon_D$. 
Taking a point in the parameter space, $m_\chi = 2$~GeV, $\kappa^2 = 10^{-7}$, $m_V = 300$~MeV, $\alpha_D = 0.5$ and the energy of 
incoming proton beam of 400~GeV, we estimate a probability of producing a $\Upsilon_D$ decaying to $3(\mu^+\mu^-)$ for a 1 mm tungsten target, $P= n \sigma \ell \sim 2\times 10^{-17}$. 
With $O(10^{20})$ particles on target, one could potentially expect up to $2\times10^{3}$ six muon events. The large multiplicity of signal events gives some hope that this signal 
could be extracted from large number of muons produced per each proton spill. Given the current uncertainties in estimating the background, we refrain from showing the 
potential reach of proton experiments in Fig.~\ref{Fig:DP-BABAR}, noting that in any case, it would not cover the most interesting region for SIDM, namely 
$m_V < 200$~MeV.

\smallskip
\noindent{\sf\bfseries Outlook.}
Among the various probes of dark sectors suggested and conducted in recent years, only a few are 
sensitive to both the dark force and dark matter at the same time. We have pointed out that in case of relatively strong self-interaction,
the presence of dark force greatly facilitates the discovery of the entire sector, as it leads to the formation of dark bound states, and causes 
dark FSR radiation that decay into multiple charged particles of the SM. The existing searches at \babar\ and Belle already limit this possibility, and further 
advance in sensitivity can be made by searching for the missing energy plus pairs of charged particles. 

\smallskip
\noindent{\sf\bfseries Acknowledgement.}
We would like to thank Clifford Cheung, Ying Fan, Ming Liu, Mark Wise and Hai-bo Yu for useful discussions.
We thank Yang Bai and Susanne Westhoff for pointing out an overall factor of (1/3) in the decay rate Eq.~\eqref{decayrate1} which was missed in the previous version.
H.A. is supported by the Walter Burke Institute at Caltech and by DOE Grant de-sc0011632.
B.E. is supported by the U.S. Department of Energy (DoE) under grant DE-FG02-92ER40701 and DE-SC0011925.
M.P. is supported in part by NSERC,
Canada, and research at the Perimeter Institute is supported
in part by the Government of Canada through
NSERC and by the Province of Ontario through MEDT.
Y.Z. is supported by the Gordon and Betty Moore Foundation through Grant \#776 to the Caltech Moore Center for Theoretical Cosmology and Physics, and by the DOE Grant DE-FG02-92ER40701, and also by a DOE Early Career Award under Grant No. DE-SC0010255.
H.A., M.P. and Y.Z. acknowledge the hospitality from the Aspen Center for Physics and the support from NSF Grant \#PHY-1066293.

\end{document}